\documentclass[12pt,prb,aps]{revtex4-1}
\usepackage{graphicx}
\usepackage{rotating}
\usepackage{array}
\usepackage{amsmath}
\usepackage{multirow}
\usepackage{setspace}
\usepackage{soul}
\usepackage{color}
\usepackage{xcolor}

\begin{document}
\title{Coupled-cluster treatment of molecular strong-field ionization}
\author{Thomas-C. Jagau\\
{\small Department of Chemistry, University of Munich (LMU), 
D-81377 Munich, Germany}}

\begin{abstract}
Ionization rates and Stark shifts of H$_2$, CO, O$_2$, H$_2$O, and CH$_4$ in static electric fields 
have been computed with coupled-cluster methods in a basis set of atom-centered Gaussian functions
with complex-scaled exponent. Consideration of electron correlation is found to be of great importance 
even for a qualitatively correct description of the dependence of ionization rates and Stark shifts on the 
strength and orientation of the external field. The analysis of the second moments of the molecular 
charge distribution suggests a simple criterion for distinguishing tunnel and barrier suppression ionization 
in polyatomic molecules. 
\end{abstract}

\maketitle

\clearpage

\section{Introduction} \label{sec:in}
Molecules exposed to electric or electromagnetic fields of a strength comparable to the internal 
molecular forces undergo ionization, possibly accompanied by dissociation.\cite{scrinzi06,lein07} 
This process underlies numerous phenomena involving strong fields such as molecular high 
harmonic generation,\cite{marangos16} laser-induced electron diffraction,\cite{zuo96} and Coulomb 
explosion.\cite{vager89} Therefore, the quantitative modeling of molecular strong-field ionization 
rates is of immediate interest for the interpretation of all experiments in which strong fields are 
applied. 

At low values of Keldysh's adiabaticity parameter,\cite{keldysh65} that is, at low frequencies and 
high intensities, the quasistatic approximation is valid: If the external field varies slowly compared 
to the inherent time scale of the ionization process, the molecule behaves at every instant as if it 
was exposed to a static field of the current strength. Ionization occurs because electrons can tunnel
through the potential barrier formed by the molecular potential and the external field or at even 
higher field strengths leave above the barrier. Differences between static and time-dependent fields 
can be treated in terms of perturbation theory in the low-frequency limit.\cite{martiskainen17} 

Within the quasistatic approximation, the ionization process can be modeled based on static-field 
ionization rates, but their computation is beyond the reach of quantum-chemical methods for bound 
states because the interaction with the field turns all bound states into Stark resonances.\cite{
nhqmbook} This is not of practical importance if the external field is weak compared to the internal 
forces and the response of the molecule can be treated in terms of perturbation theory.\cite{
helgaker12} However, a perturbative approach is invalid in the quasistatic regime; in Hermitian 
quantum mechanics, the static-field ionization rate can only be determined from the time-dependent 
Schr\"odinger equation.\cite{parker09}

On the contrary, a time-independent treatment is possible in non-Hermitian quantum mechanics.\cite{
nhqmbook} The ionization rate $\Gamma$ induced by the external static field $F$ is associated 
with the imaginary part of a discrete complex eigenvalue 
\begin{equation} \label{eq:cenergy}
E - i \, \Gamma/2
\end{equation}
of the molecular Hamiltonian. Similarly, the Stark shift $\Delta E$ can be calculated by comparing 
the real part of the resonance energy to the field-free case. However, since the Stark resonances 
are not $L^2$ integrable, they cannot be treated using quantum chemistry for bound states. 

In the case of atomic Stark resonances, complex scaling \cite{aguilar71,balslev71} is a handy 
solution even though the electric field is not dilation analytic. Upon scaling all coordinates in the 
Hamiltonian by a complex number $e^{i\theta}$, the Stark resonance wave function becomes 
$L^2$ integrable provided that $\theta$ exceeds a critical value.\cite{herbst78,herbst79,herbst81,
nicolaides92} The eigenvalues of the complex-scaled Hamiltonian can then be computed in 
analogy to bound states and are interpreted according to Eq. \eqref{eq:cenergy}.\cite{reinhardt76,
nicolaides93,scrinzi99,jagau16b}

For a molecule in the Born-Oppenheimer approximation, complex scaling is not appropriate.\cite{
nhqmbook} Several alternative complex-variable (CV) techniques have been proposed, notably 
exterior complex scaling,\cite{simon79} complex scaling of the Hamiltonian's matrix elements,\cite{
moiseyev79} the use of a complex-scaled basis,\cite{mccurdy78} complex-absorbing potentials 
(CAPs),\cite{riss93} and reflection-free CAPs.\cite{moiseyev98} CAPs have evolved to the most 
popular CV technique for autoionizing resonances (see Ref. \citenum{resrev} for an overview of 
recent work) and have also been used to investigate Stark resonances induced by time-dependent 
fields with explicitly time-dependent configuration-interaction singles (TD-CIS).\cite{krause14,
krause15a,krause15b} On the contrary, electronic-structure calculations in a basis of complex-scaled 
functions have been carried out only for autoionizing resonances.\cite{honigmann06,white15a,white15b,
white17}

The computation of molecular static-field ionization rates is a topic of current research; important 
recent contributions rely on the hybrid antisymmetrized coupled-channels (haCC) approach \cite{
majety15a,majety15b} or the weak-field asymptotic theory \cite{tolstikhin14,madsen14,walt15,
yue17} or apply more drastic approximations such as the popular formula by Ammosov, Delone, 
and Krainov \cite{ammosov86} and its extensions.\cite{tong02,tong05} 

In this work, the method of complex basis functions is applied to molecular Stark resonances 
induced by static electric fields. The many-body electronic Schr\"odinger equation is solved 
within the coupled-cluster singles and doubles (CCSD) approximation \cite{purvis82} and 
within the CCSD approximation with additional perturbative triples excitations (CCSD(T)).\cite{
raghavachari89,bartlett90} The definitions of these methods are the same for all CV techniques 
and identical to standard CC theory \cite{ccbook} apart from the different metric owing to the 
non-Hermiticity of the Hamiltonian.\cite{resrev} 

A particular advantage of a CC treatment of molecular Stark resonances is that all ionization channels 
can be computed as eigenstates of the same Hamiltonian in a biorthogonal representation through 
the equation-of-motion (EOM) CC formalism.\cite{eomrev1,eomrev2} Thus, their characterization 
through Dyson orbitals is straightforward.\cite{oana07,jagau16a} Also, the CC formalism for molecular 
properties can be applied to compute moments of the electronic charge distribution, which provides 
further insight into the ionization process. 

The article is organized as follows: Section \ref{sec:meth} covers technical aspects of the 
computational scheme, while Section \ref{sec:two} discusses several conceptual aspects of 
molecular Stark resonances by means of the two-electron systems He and H$_2$, for which 
CCSD is equivalent to full configuration interaction (CI). Section \ref{sec:many} presents 
representative applications to CO, O$_2$, H$_2$O, and CH$_4$ and Section \ref{sec:conc} 
provides some conclusions and an outlook.

\section{Computational Details} \label{sec:meth}  
All calculations have been carried out with a development version of the Q-Chem program package, 
release 5.0.\cite{qchem} The implementation of Gaussian basis functions with a complex-scaled 
exponent presented in Ref. \citenum{white15a} is reused. The only piece of code additionally required 
for Stark resonances is the evaluation of the dipole integrals in the complex-scaled basis, for which 
the same formulas apply as in the case of real algebra.\cite{matsuoka71} The implementation of 
Hartree-Fock (HF) theory in a complex-scaled basis described in Ref. \citenum{white15b} has also 
been reused with some modifications required due to the differences in the spectrum of the field-including 
Hamiltonian and its field-free counterpart.\cite{jagau16b} Technical aspects of CV-CC methods have 
been discussed for temporary anions in Refs. \citenum{bravaya13,zuev14} and apply equally to Stark 
resonances. 

Similar basis sets as those suggested for temporary anions \cite{white15a} are employed in all 
calculations. These bases consist of an unscaled part, which is always chosen as cc-pVQZ 
in this work, and additional diffuse functions at all atoms whose exponents are subject to complex 
scaling. The latter functions are chosen as the standard diffuse functions from aug-cc-pVQZ plus 
additional even-tempered s, p, d, and f functions. Details about the basis sets are compiled in the 
supplementary material. Overall sizes range from 188 functions for H$_2$ up to 312 functions for 
CH$_4$. 

In the employed basis sets, the imaginary part of the field-free energy can grow as large as 
$\sim$0.001 a.u. at some values of the scaling angle $\theta$ similar to what has been observed 
in CC calculations of atomic Stark resonances using a complex-scaled Hamiltonian.\cite{jagau16b} 
Therefore, it is essential to correct the resonance energies according to 
\begin{equation} \label{eq:corr}
E'_\text{res} (F, \theta) = E_\text{res}(F, \theta) - E_\text{res}(F=0, \theta) + E_\text{res}(F=0,\theta=0)
\end{equation}
as proposed in Ref. \citenum{jagau16b} and $\Delta E$ and $\Gamma$ are then evaluated from $E'$ 
at $\theta_\text{opt} = \theta \, \Big\vert\, \text{min} \big( |dE'_\text{res}(\theta)/d\theta| 
\big)$.\cite{moiseyev78} The dependence of $\theta_\text{opt}$ on the quantum-chemical method 
is weak in most cases so that $\theta_\text{opt}$ from a HF calculation can be used as guess in 
a subsequent CC calculation. This even holds in cases where HF and CC calculations yield $\Gamma$ 
values differing by an order of magnitude. On the contrary, different molecules and field strengths 
can feature significantly different $\theta_\text{opt}$. Typical values of $\theta_\text{opt}$ are 
in the range of 10--25$^\circ$. In this work, $\theta_\text{opt}$ has been determined to $1^\circ$, 
which is sufficient to evaluate $\Delta E$ and $\Gamma$ with a relative precision of $<1\%$ for all 
cases considered. 

\section{Results for two-electron systems}
\label{sec:two}
\subsection{Complex-scaled basis functions vs. complex-scaled Hamiltonian}
\label{sec:cs}
Atomic Stark resonances can be treated by complex scaling. For helium an expansion of the 
complex-scaled Hamiltonian in the aug-cc-pVQZ basis with additional diffuse s, p, d, and f 
functions yields ionization rates \cite{jagau16b} that agree within 1-2\% with reference data 
obtained by integrating the time-dependent Schr\"odinger equation \cite{parker09} or from 
complex-scaled calculations in a basis of explicitly-correlated two-electron functions.\cite{scrinzi99}  

Representing the Hamiltonian in a partially complex-scaled basis as detailed in Section 
\ref{sec:meth} constitutes an approximation to exterior complex scaling. As documented 
in Table S2 in the Supplementary Material, ionization rates for helium obtained with the 
partially complex-scaled basis consistently overestimate the reference data from explicit 
complex scaling by about 5\% for a wide range of field strengths, while Stark shifts deviate 
by less than 1\%. Considering the great sensitivity of $\Gamma$ towards truncation of the 
one-particle basis set and approximations to the many-body treatment,\cite{jagau16b} a 
consistent deviation of 5\% appears entirely acceptable. 

Notably, at very high ($F > 0.45$ a.u.) and at low field strengths ($F < 0.14$ a.u.) the 
deviations between $\Gamma$ values become significantly larger. In the low-field limit, this 
is because $\Gamma$ itself becomes very small and its dependence on the scaling angle 
$\theta$ masks the effect of the field. Whereas complex scaling appears to be applicable 
to ionization rates as low as $\sim 10^{-6}$ a.u., \cite{jagau16b} reliable calculations in a 
complex-scaled basis require somewhat larger $\Gamma$ values of at least $\sim 10^{-5}$ 
a.u. In the high-field limit, better agreement with explicit complex scaling is observed if more 
functions in the basis set are scaled. Importantly, however, scaling more basis functions than 
proposed in Ref. \citenum{white15a} for temporary anions does not reduce the 5\% deviation from 
explicit complex scaling in general.

\subsection{Potential energy curve of H$_2$}
\label{sec:h2}
\begin{figure}[htbp]
\includegraphics[scale=0.895]{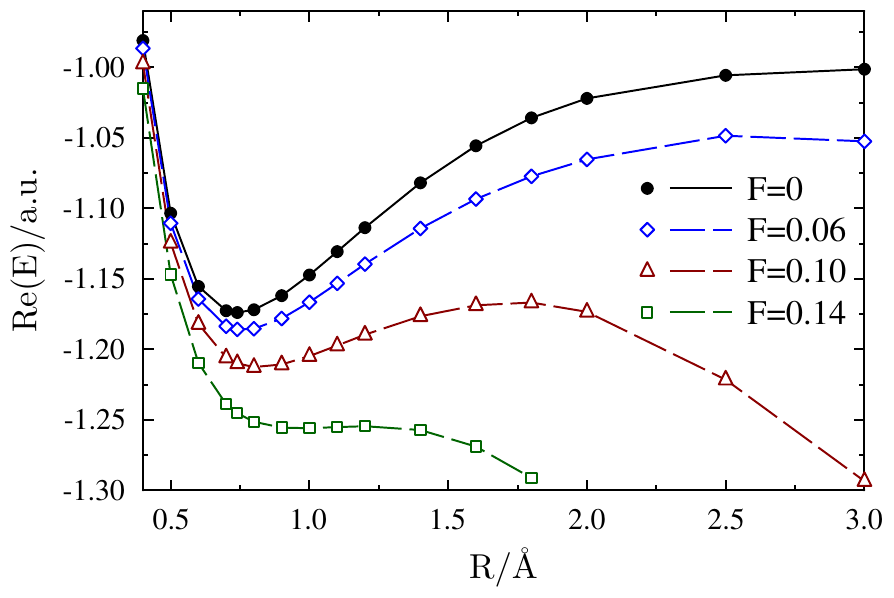}
\includegraphics[scale=0.895]{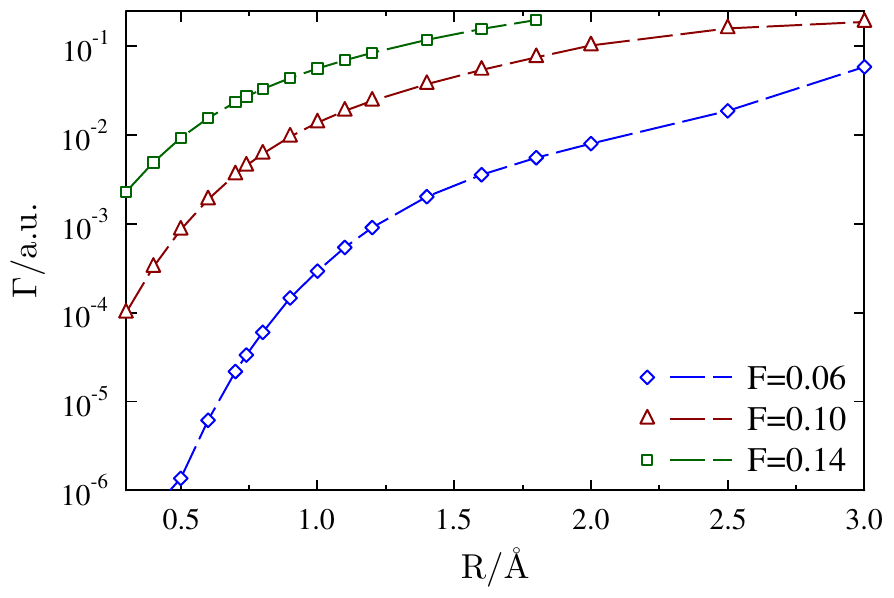}
\includegraphics[scale=0.895]{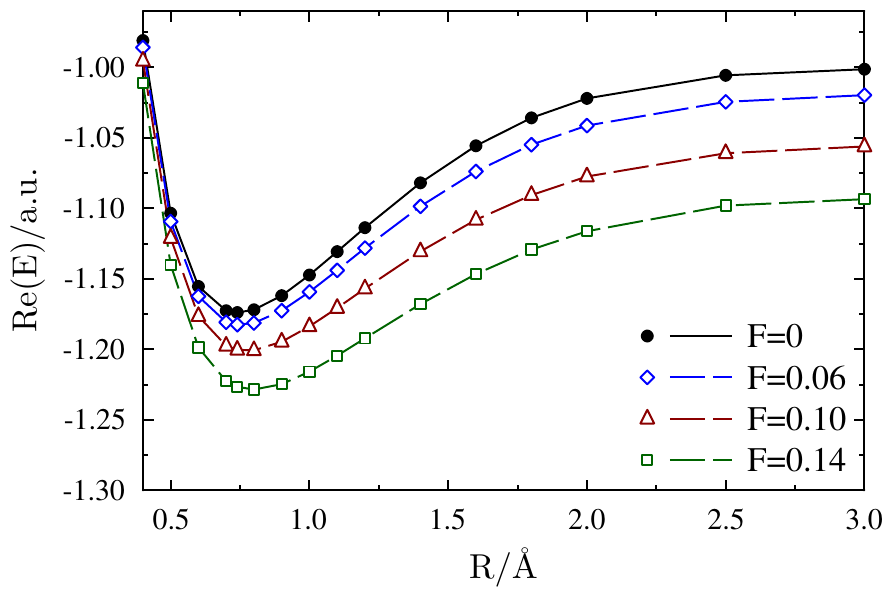}
\includegraphics[scale=0.895]{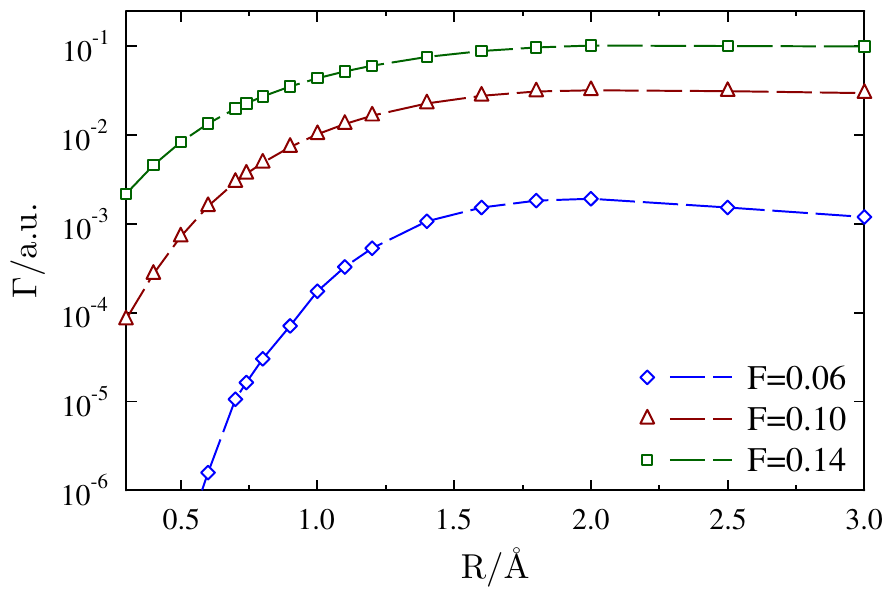}
\caption{Potential energy curve (left) and ionization rate (right) of H$_2$ at different field strengths 
oriented parallel (upper panels) or perpendicular (lower panels) to the molecular axis computed 
at the full CI level of theory using a modified aug-cc-pVQZ basis set.}
\label{fig:h2} \end{figure}

The electronic Schr\"odinger equation for H$_2$ can be solved exactly within a given 
basis set for arbitrary HH distances and orientations of the external field with the 
present implementation. Therefore, H$_2$ is the natural first molecular test case.
Several representative potential energy curves (PECs) are compiled in Figure \ref{fig:h2}. 
The real parts of such complex-valued PECs can be interpreted in analogy to bound 
states, while the imaginary part yields the ionization rate as a function of the molecular 
structure.\cite{moiseyev17} The case with the field being parallel to the molecular axis 
(upper two panels of Figure \ref{fig:h2}) has been considered previously \cite{saenz00,
saenz02} with the Hamiltonian expressed in a basis of explicitly correlated two-electron 
functions. Ionization rates from this approach agree within 3\% with the present approach 
for $F$ = 0.06--0.14 a.u. and R(HH)=0.74 \AA\ (see SI for details). 

As Figure \ref{fig:h2} illustrates, full CI calculations in a complex-scaled basis set produce 
smoothly varying energies and ionization rates, at least at the level of precision considered 
here with R(HH) varied in steps of 0.1 \AA. Discontinuities may appear upon zooming in 
because the optimal scaling angle $\theta_\text{opt}$ varies as a function of the molecular 
structure. However, this effect seems to be smaller for Stark resonances than for autoionizing 
resonances, likely because the perturbation of the field-free state by complex scaling 
can be removed according to Eq. \eqref{eq:corr} for Stark resonances. 

Figure \ref{fig:h2} also shows that already a field strength of 0.14 a.u. is sufficient to make the 
minimum in the PEC disappear in the case of a parallel field. Whereas the PEC will always 
acquire dissociative character eventually if there is a component of the field along the molecular 
axis, this does not happen if the field is exactly perpendicular. In this latter case, the PEC is just 
Stark shifted, but retains its field-free shape. In the limit R(HH)$\to\infty$ the Stark shift and the 
ionization rate converge to twice the values of the hydrogen atom. Interestingly, $\Gamma$ 
approaches that limit from above at $F=0.06$ a.u. and exhibits a maximum at around R(HH) 
= 2.0 \AA. This maximum is, however, not to be confused with the maximum in the diabatic 
ionization rate that occurs at around 2.8 \AA\ when the field is parallel to the molecular axis 
and is caused by an avoided crossing of two PECs.\cite{saenz00,saenz02,seideman95,zuo95} 
Finally, Figure \ref{fig:h2} also illustrates that $\Gamma$ converges to the ionization rate of 
the He atom in the limit R(HH)$\to 0$ for both orientations of $F$.

\subsection{Below vs. above barrier ionization}
\label{sec:r2}
Depending on the strength of the external field, tunnel and barrier suppression ionization can be 
distinguished. For atoms, the field strength where the transition between the two regimes occurs 
can be estimated as 
\begin{equation} \label{eq:fabi}
F_\text{ABI} = I_p^2/4
\end{equation}
with $I_p$ as the lowest ionization potential. In molecular strong-field ionization, the same distinction 
is possible, but there is no simple estimate of the critical field strength akin to Eq. \eqref{eq:fabi}. To 
characterize ionization of diatomic molecules with one electron by a field oriented along the molecular 
axis, a double-well model can be used,\cite{bandrauk12} but a generalization to polyatomic molecules 
with more complicated nuclear configurations and several competing ionization channels does not 
appear to be straightforward. 

Recently, it was demonstrated for various atoms that the transition from tunnel to barrier suppression 
ionization is accompanied by a marked change in the second moment of the electronic charge 
distribution, i.e., the spatial extent of the wave function in the direction of the external field.\cite{
jagau16b} While the resonance wave function becomes more extended with increasing field strength 
in the tunnel ionization regime, the opposite trend is observed if the ionization takes place above 
the barrier. The maximum in $\langle z^2 \rangle (F)$ ($F$ parallel to $z$-axis) coincides very well 
with the estimate from Eq. \eqref{eq:fabi}. \cite{jagau16b}

\begin{figure}[htbp]
\includegraphics[scale=0.895]{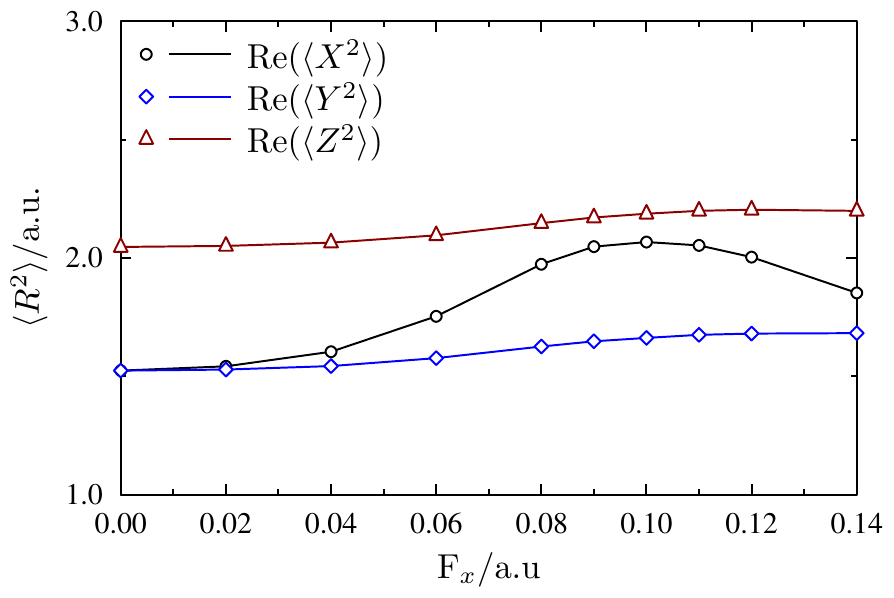}
\includegraphics[scale=0.895]{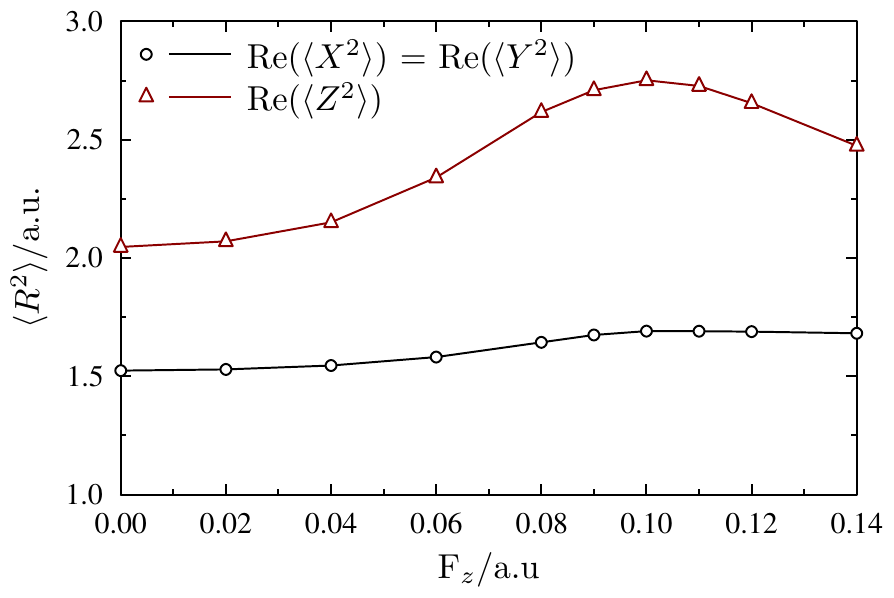}\\
\includegraphics[scale=0.895]{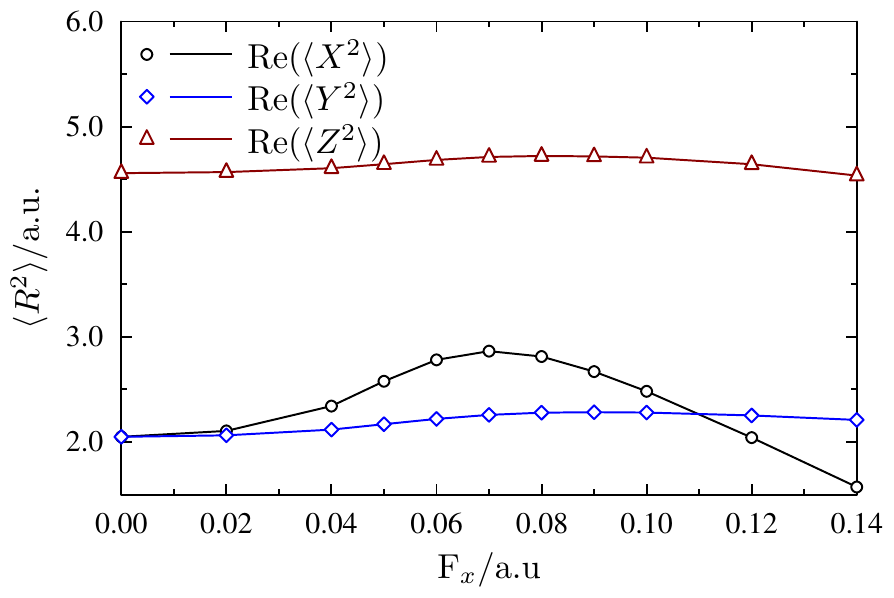}
\includegraphics[scale=0.895]{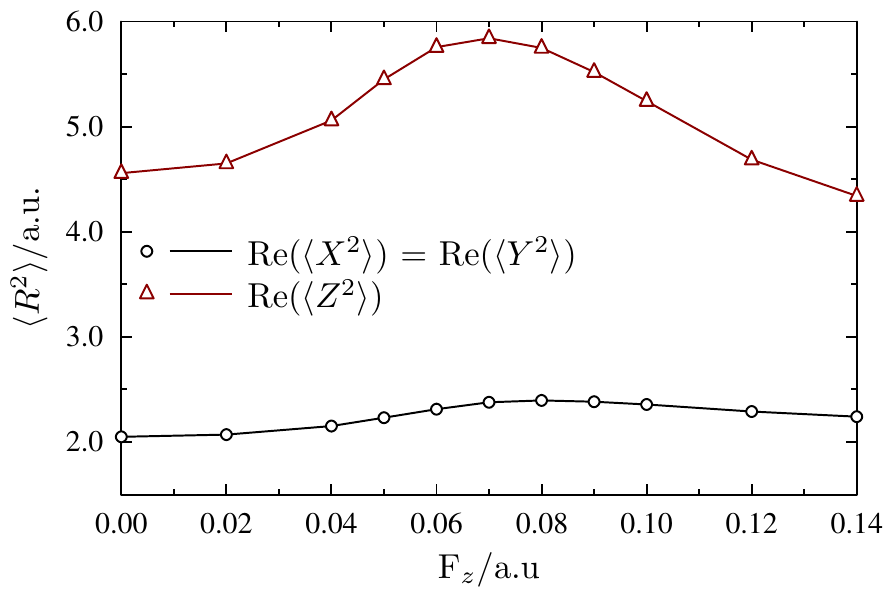}
\caption{Real parts of second moments $\langle R^2 \rangle$ for H$_2$ computed at the full CI level 
of theory using a modified aug-cc-pVQZ basis set. The field is oriented either perpendicular (left panels) 
or parallel (right panels) to the molecular axis (=$z$-axis). The HH distance is 0.74 \AA\ (upper panels) 
or 1.40 \AA\ (lower panels).}
\label{fig:h2p} \end{figure}

Figure \ref{fig:h2p} illustrates that similar trends in the second moment are observed for the H$_2$ 
molecule. In each of the four panels, which relate to different orientations of the external field 
and HH bond lengths, a distinct maximum is observed in the component of the second moment in 
the direction of the field. If the field is perpendicular to the molecular axis (left two panels 
of Figure \ref{fig:h2p}), the potential that the outgoing electron needs to overcome is similar to the 
atomic case. Consequently, the maximum of $\langle x^2 \rangle$ agrees well with the estimate of 
$F_\text{ABI}$ from Eq. \eqref{eq:fabi} as documented in Table \ref{tab:h2fabi}. When the field is 
parallel to the molecular axis (right two panels of Figure \ref{fig:h2p}), a double-well model is appropriate, 
but in the case of a many-electron system it is unknown to what degree the charges of the two nuclei 
are screened. Assuming effective charges of 0.5 a.u. for both nuclei leads to significant deviations 
of $F_\text{ABI}$ from the maxima of $\langle z^2 \rangle$ as Table \ref{tab:h2fabi} shows. 
Better agreement is obtained if one applies the formula for the atomic case (Eq. \eqref{eq:fabi}) 
assuming that one nucleus is completely screened while the other one is completely unscreened. 

The significance of the trends in the second moment lies in the fact that this quantity can be 
easily computed for polyatomic molecules and thereby offers a clear criterion to distinguish 
tunnel and barrier suppression ionization in arbitrary many-electron systems without assumptions 
about their electronic structure. As will be shown in Section \ref{sec:many}, a characteristic 
maximum is also observed for O$_2$, H$_2$O, and CH$_4$.

\section{Results for many-electron systems}
\label{sec:many}

\subsection{Carbon monoxide}
\label{sec:co}
CO has been chosen as the first example of a many-electron system because experimental 
results are available regarding the angular dependence of the ionization rate \cite{li11,wu12} 
and an accurate treatment of the electronic structure has proven to be important for a correct 
description of the ionization process.\cite{li11,wu12,zhang13,spiewanowski15,majety15a} 
Moreover, ionization rates from the haCC approach have been reported in the literature.\cite{
majety15a} 

\begin{figure}[htbp]
\includegraphics[scale=0.895]{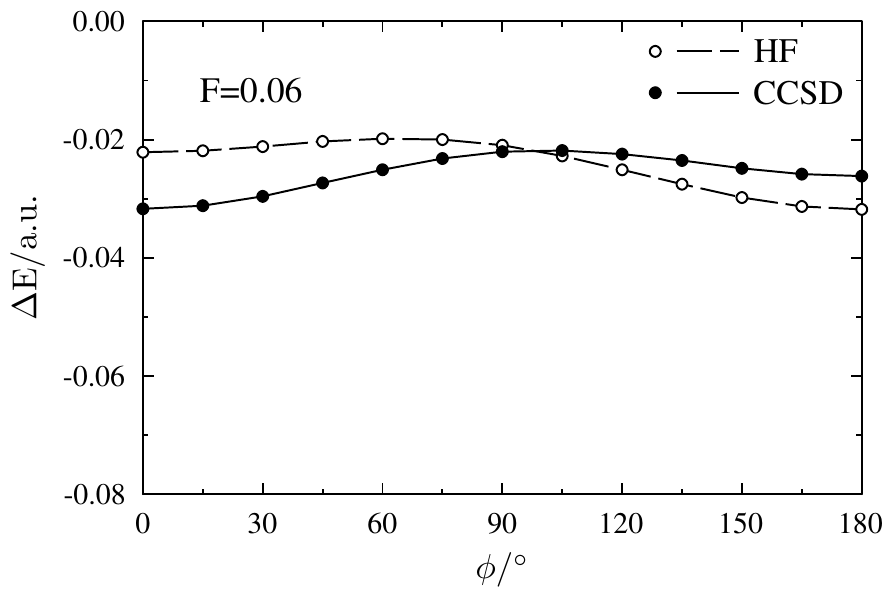}
\includegraphics[scale=0.895]{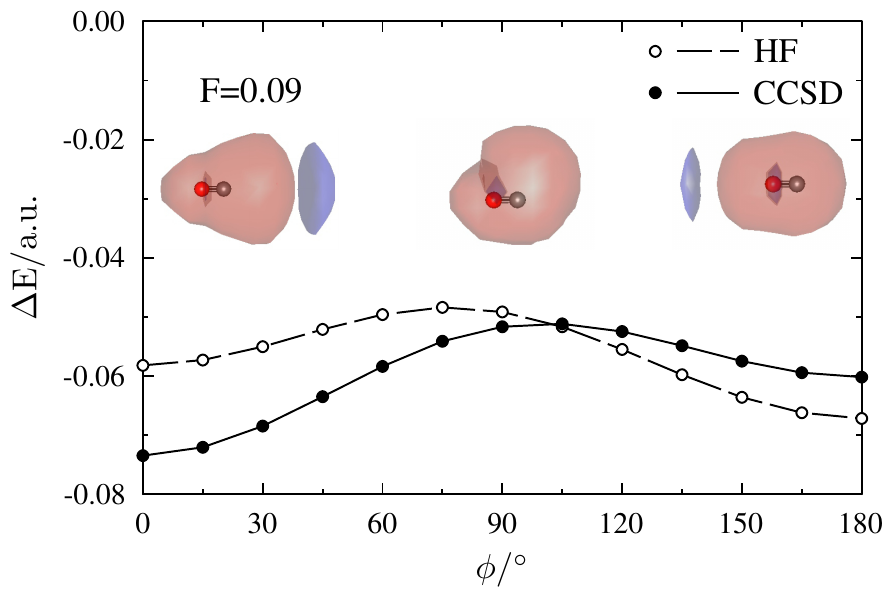}
\includegraphics[scale=0.895]{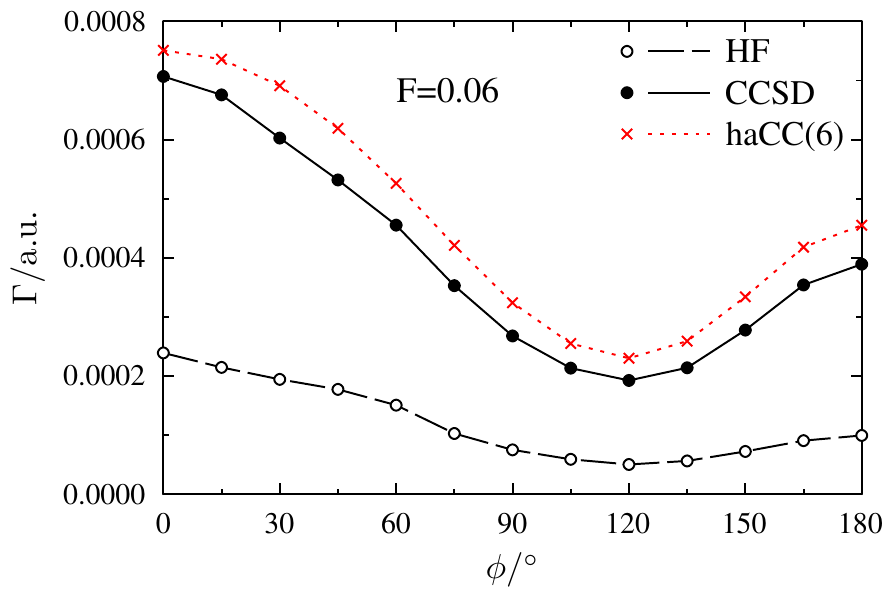}
\includegraphics[scale=0.895]{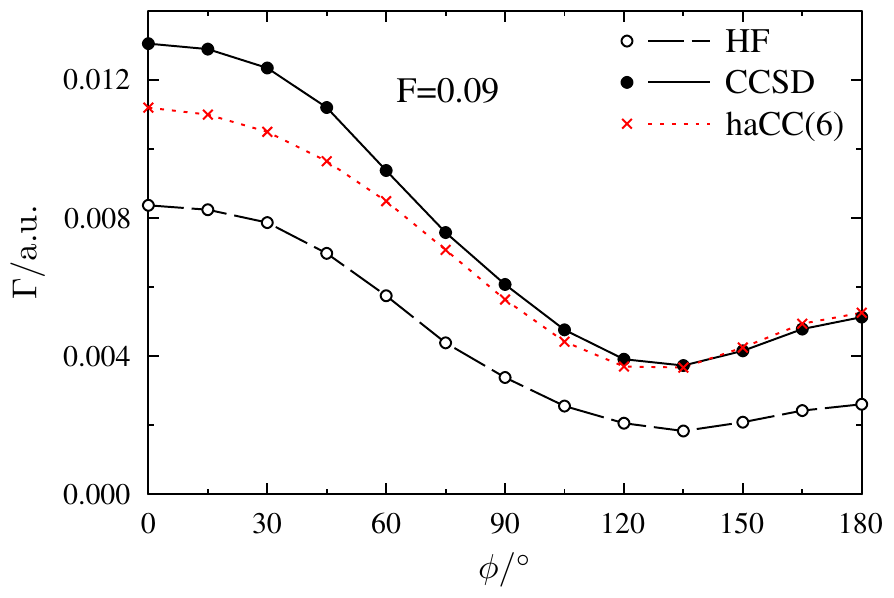}
\caption{Angle dependent Stark shifts $\Delta E$ (upper panels) and ionization rates $\Gamma$ (lower 
panels) of CO at field strengths of $F=0.06$ a.u. (left) and $F=0.09$ a.u. (right) computed at the HF and 
CCSD levels of theory using a modified aug-cc-pVQZ basis set. Ionization rates from Ref. \citenum{
majety15a} obtained with the haCC(6) approach are also shown. $\phi = 0^\circ$ corresponds to the field 
pointing from C to O. The real part of the Dyson orbital for decay into the ground state of CO$^+$ is shown 
as inset at 0$^\circ$, 90$^\circ$, and 180$^\circ$.}
\label{fig:co} \end{figure}

Figure \ref{fig:co} shows Stark shifts and ionization rates computed with HF and CCSD at 
$F=0.06$ a.u. and $F=0.09$ a.u. as a function of the angle between the external field and 
the molecular axis. The upper panels illustrate that HF and CCSD disagree whether the 
Stark shift is larger when the field points towards the carbon atom or towards the oxygen 
atom. This is not surprising given that electron correlation reverses the sign of the dipole 
moment of CO. On the contrary, HF and CCSD qualitatively agree about the angular 
dependence of the ionization rate: $\Gamma$ is higher when the electron leaves towards 
the C atom than when it leaves towards the O atom consistent with experimental findings.\cite{
li11,wu12} The minimum in $\Gamma$ occurs at an angle of about 120$^\circ$ ($F=0.06$ 
a.u.) or 135$^\circ$ ($F=0.09$ a.u.) with both methods, but HF underestimates the absolute 
value of $\Gamma$ by a factor of 3 at $F=0.06$ a.u. and still by a factor of 1.5 at $F=0.09$ 
a.u., which is similar to trends in atomic Stark resonances.\cite{jagau16b} The anisotropy 
of the ionization rate is underestimated at $F=0.06$ a.u. within the HF approximation but in 
qualitative agreement with CCSD at $F=0.09$ a.u.

Carbon monoxide has two low-lying ionized states: a $^2\Sigma$ state and a $^2\Pi$ state 
whose energies are obtained with EOM-IP-CCSD at $F=0$ as 14.23 eV and 17.16 eV, respectively. 
The relative energy of the $^2\Sigma$ state rises or falls by more than 2 eV depending on the 
orientation of the external field at $F=0.09$ a.u., whereas the $^2\Pi$ state moves by only 
0.3 eV at the same field strength. Although the computation of partial widths for decay into 
specific ionization channels is beyond the present work, their Dyson orbitals provide qualitative 
insight. Specifically, only the Dyson orbital for the lowest ionized state (depicted as inset 
in Figure \ref{fig:co}) acquires a non-negligible imaginary part at the field strengths 
considered here, which indicates that decay into this channel dominates the ionization process. 
The Dyson orbital at 90$^\circ$ also illustrates that the lowest ionized state largely retains 
its $\Sigma$ character even though the external field breaks spatial symmetry and $\Sigma$ and 
$\Pi$ can mix.

Figure \ref{fig:co} also shows ionization rates obtained with the haCC approach with 6 cationic 
states included in the wave function of the resonance. At $F=0.06$ a.u., CCSD and haCC(6) 
qualitatively agree about the ionization rate and its anisotropy, but haCC(6) yields $\Gamma$ 
values that are consistently larger (4--8 $\cdot$ $10^{-4}$ a.u., 7--15\%). At $F=0.09$ a.u., CCSD 
and haCC(6) agree very well ($1 \cdot 10^{-4}$ a.u., 2--3\%) when the electron leaves towards 
the O atom (180$^\circ$), but the deviation grows up to 0.002 a.u. (16 \%) at 0$^\circ$ so that 
CCSD predicts a considerably higher anisotropy of $\Gamma$ than haCC(6). 

The discrepancy at $F=0.09$ a.u. may reflect the systematic underestimation of ionization 
rates at high field strengths in haCC due to the ionization channels being treated as 
bound states.\cite{majety15a} The origin of the discrepancy at $F=0.06$ a.u. is less clear 
and could indicate a shortcoming of the present approach, for example, an insufficient 
one-electron basis or non-negligible electron correlation beyond the CCSD approximation. 
Additional calculations at $\phi=0^\circ$ and $\phi=180^\circ$ ($F=0.06$ a.u.) including 
further diffuse basis functions change $\Gamma$ by only $\sim 10^{-5}$ a.u., whereas the 
(T) correction increases $\Gamma$ by about 20\% and thus overcompensates the difference 
between CCSD and haCC(6). Noteworthy, going from the quadruple-$\zeta$ basis to a smaller 
triple-$\zeta$ basis as used for the haCC(6) calculations in Ref. \citenum{majety15a} also 
leads to a 20\% increase of $\Gamma$.

\subsection{Dioxygen}
\label{sec:o2}
\begin{figure}[htbp]
\includegraphics[scale=0.895]{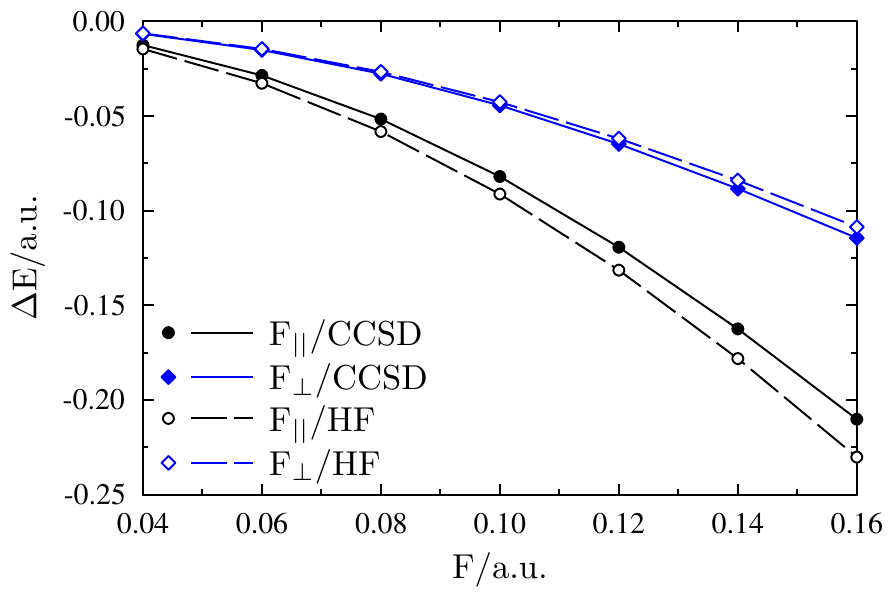}
\includegraphics[scale=0.895]{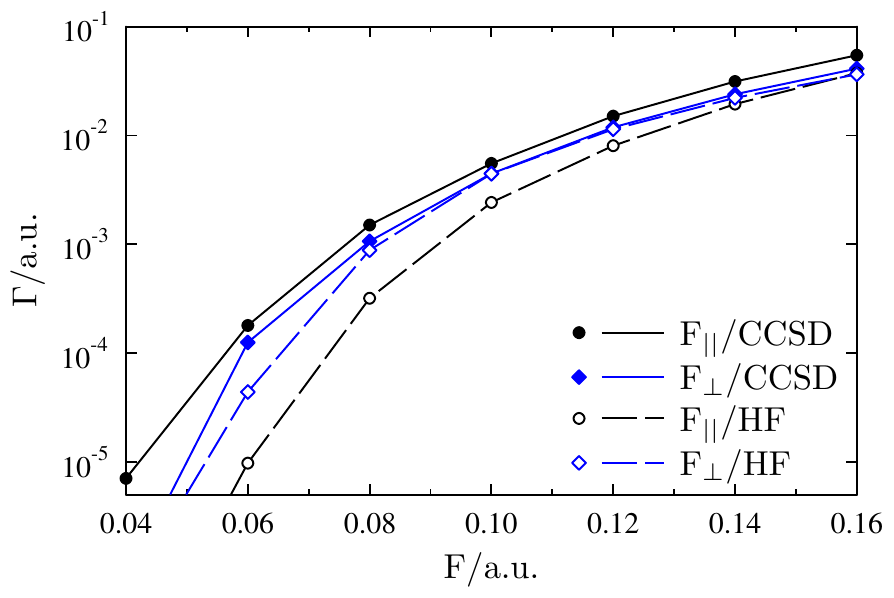}
\includegraphics[scale=0.895]{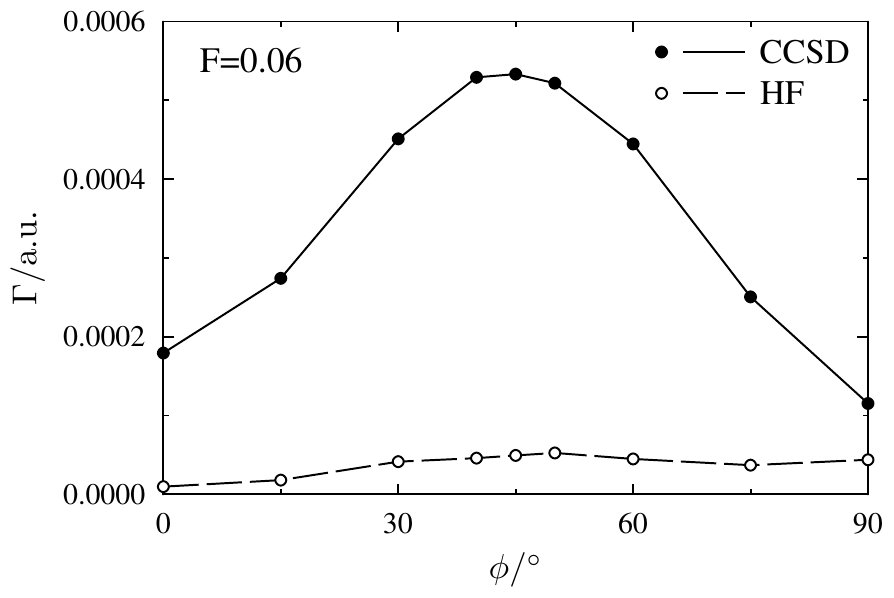}
\includegraphics[scale=0.895]{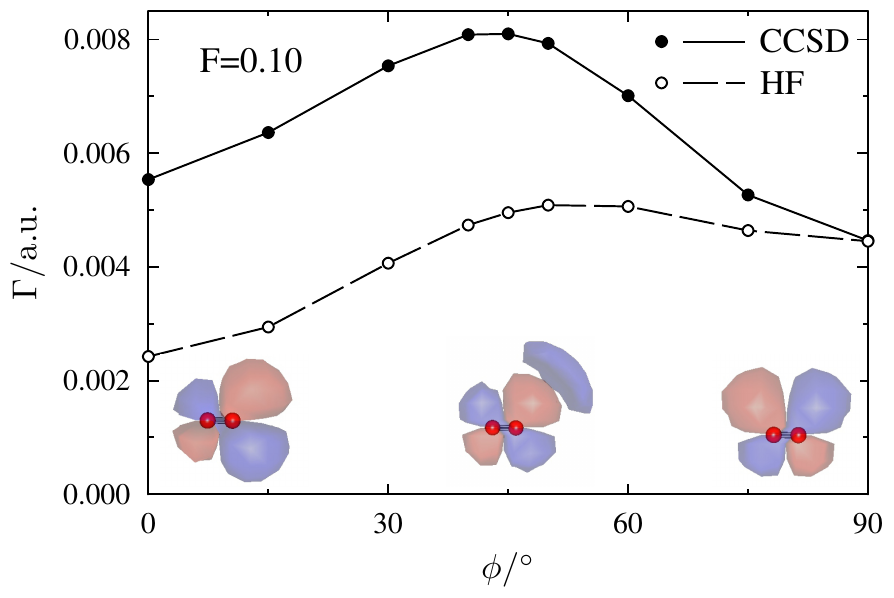}
\includegraphics[scale=0.895]{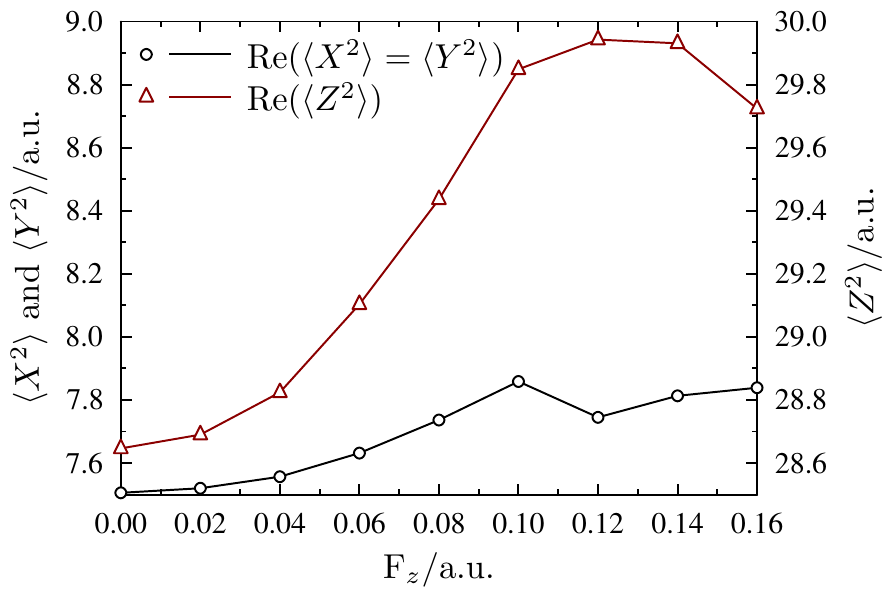}
\includegraphics[scale=0.895]{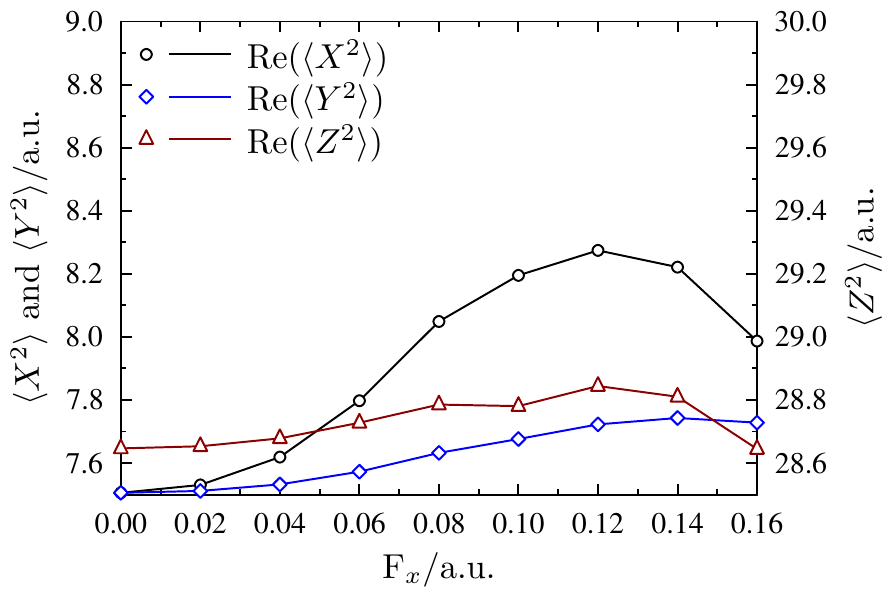}
\caption{Strong-field ionization of O$_2$ studied at the HF and CCSD levels of theory using a 
modified aug-cc-pVQZ basis set. Upper panels: Stark shifts $\Delta E$ (left) and ionization rates 
$\Gamma$ (right) as a function of field strength $F$. The field is oriented either parallel ($||$) or 
perpendicular ($\perp$) to the molecular axis (=$z$-axis). Middle panels: Angle dependent ionization 
rates at $F=0.06$ a.u. (left) and $F=0.10$ a.u. (right). $\phi = 0^\circ$ corresponds to the field 
oriented parallel to the $z$-axis. The real part of the Dyson orbital for decay into the ground state 
of O$_2^+$ is shown as inset at 0$^\circ$, 45$^\circ$, and 90$^\circ$. Lower panels: Real parts 
of the components of the second moment. The field is oriented either parallel (left panel) or 
perpendicular (right panel) to the $z$-axis.}
\label{fig:o2} \end{figure}

Stark shifts and ionization rates of O$_2$ computed with HF and CCSD are presented in 
Figure \ref{fig:o2}. Electron correlation has only minor impact on $\Delta E$ (upper left 
panel), whereas it changes $\Gamma$ dramatically. As illustrated in the upper right panel, 
CCSD finds that the ionization rate is at all field strengths higher when the field is 
oriented parallel to the molecular axis than when it is perpendicular. Also, the ratio 
$\Gamma_{||}/\Gamma_\perp \approx 1.2$--1.5 does not vary much with the field strength. 
On the contrary, at the HF level this ratio is computed to increase substantially from 
$<0.3$ to $\sim 1$ between $F=0.06$ and 0.16 a.u., i.e., HF predicts higher $\Gamma$ for 
perpendicular orientation. The discrepancy between HF and CCSD at low to medium field 
strengths is also apparent from the angle-dependent ionization rates shown in the middle 
panels of Figure \ref{fig:o2}. At $F=0.06$ a.u., HF underestimates $\Gamma$ by a factor 
of 3--20 compared to CCSD depending on the orientation, whereas that factor shrinks to 
1--2.5 at $F=0.10$ a.u. The impact of electron correlation is at all field strengths 
largest when the field is parallel to the molecular axis. The maximum ionization rate 
is obtained with CCSD at an angle of 45$^\circ$ at $F=0.06$ a.u. and 0.10 a.u. consistent 
with experimental findings,\cite{pavicic07} while HF locates the peak in $\Gamma$ at 
around 50$^\circ$. 

The lowest-lying state of O$_2^+$ ($^2\Pi_g$) is computed by EOM-IP-CCSD to lie at 12.37 
eV in the field-free case. Similar to CO, only the Dyson orbital of this state acquires 
a substantial imaginary part in the presence of the field indicating that formation of 
this state is the preferred ionization pathway. The relative energy of the $^2\Pi_g$ state 
changes by 0.5--0.9 eV at $F=0.10$ a.u. depending on the orientation and hence significantly 
less than the ground state of CO$^+$, which is expected given that O$_2$ is nonpolar. It is 
also noteworthy that the c-norm of the Dyson orbital stays close to 1 at all field strengths 
and orientations even though HF and CCSD disagree so strongly about $\Gamma$. 

The lower panels of Figure \ref{fig:o2} illustrate the dependence of the second moment of the 
electronic charge distribution on the field strength. For both parallel and perpendicular 
orientation of the external field, a characteristic maximum of one component is observed as 
discussed in Section \ref{sec:r2}. However, other than for H$_2$, the field strength where 
that maximum occurs (ca. 0.12 a.u. for both orientations) does not agree with the atomic 
estimate for the transition between tunnel and above-barrier ionization. Eq. \eqref{eq:fabi} 
yields $F_\text{ABI} = 0.056$ a.u. for perpendicular orientation, which is clearly wrong 
given the behavior of $\Gamma$ at this field strength and illustrates that Eq. \eqref{eq:fabi} 
is inapplicable to molecules with many electrons. A value of 0.12 a.u. as suggested by the 
second moment appears as a better estimate indicating that the tunnel ionization regime extends 
to a higher field strength than in a hypothetical atom with the same ionization potential. 

\subsection{Water}
\label{sec:h2o}
\begin{figure}[htbp]
\includegraphics[scale=0.895]{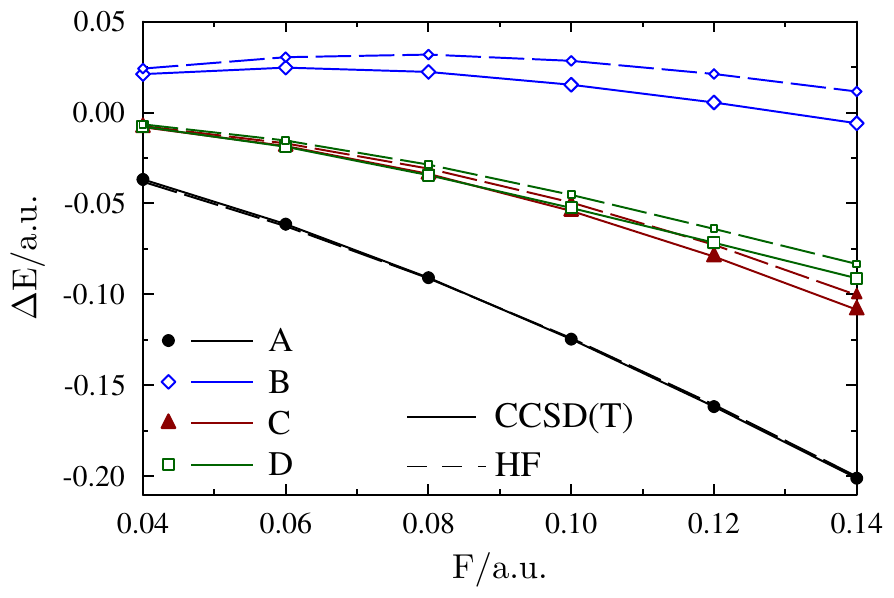} 
\includegraphics[scale=0.895]{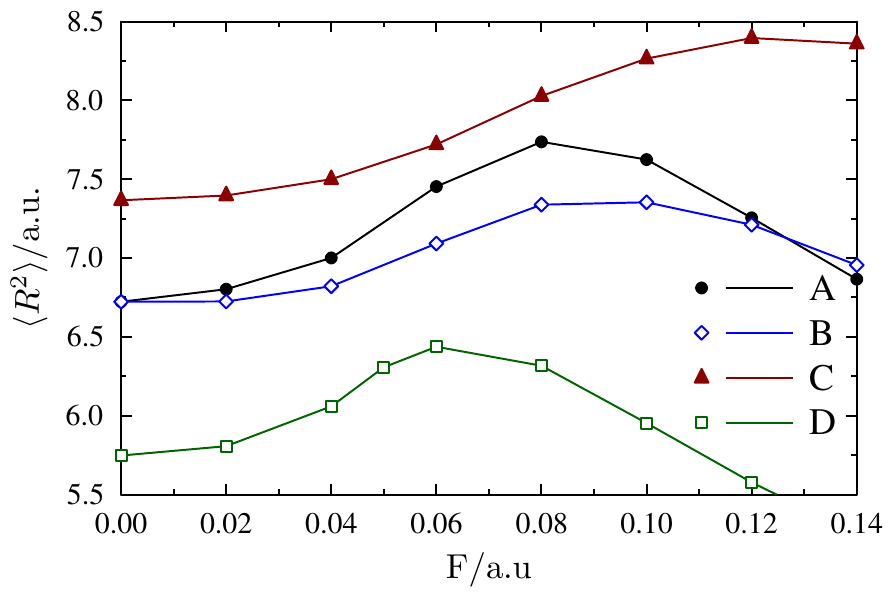}
\includegraphics[scale=0.895]{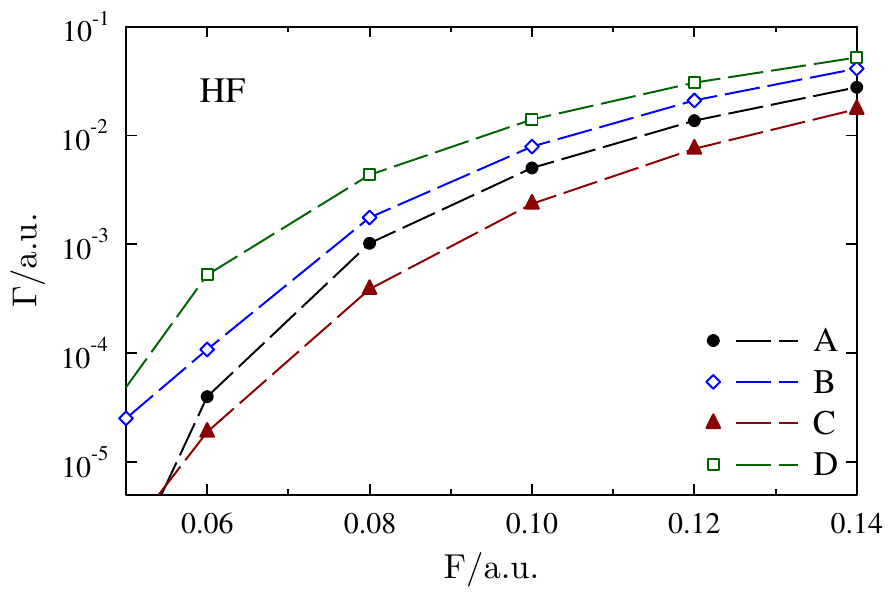}
\includegraphics[scale=0.895]{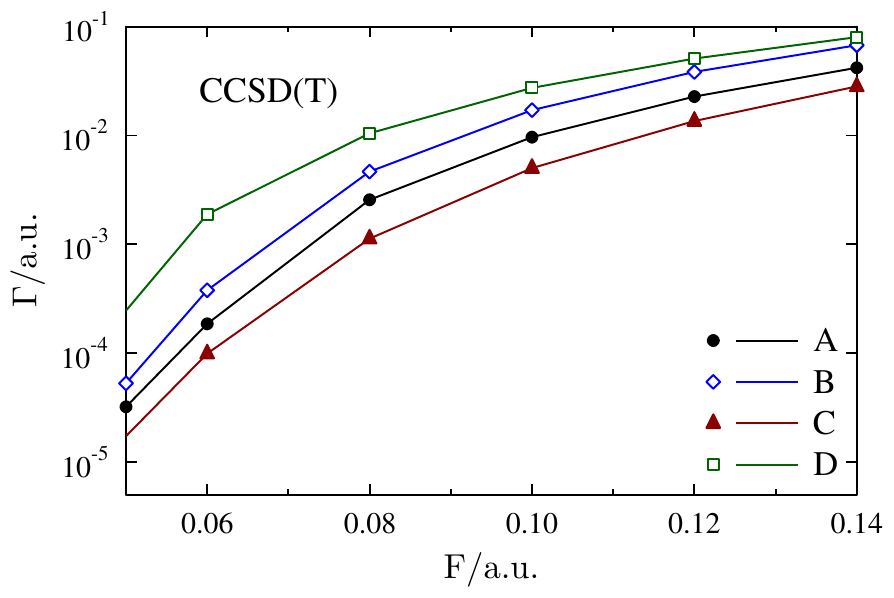} \\
\caption{Strong-field ionization of H$_2$O studied at the HF, CCSD, and CCSD(T) levels of theory 
using a modified aug-cc-pVQZ basis set. Upper left panel: Stark shifts $\Delta E$ as a function of 
field strength. Upper right panel: Real part of the component of the second moment in the direction 
of the external field as a function of field strength. Lower panels: Ionization rates $\Gamma$ as a 
function of field strength. The field is oriented as follows: \\ 
A --- along the molecular axis (=$z$-axis), away from the oxygen atom, \\
B --- along the molecular axis (=$z$-axis), towards the oxygen atom,\\
C --- perpendicular to the molecular axis in the molecular plane (= along the $y$-axis),\\
D --- perpendicular to the molecular plane (= along the $x$-axis)}
\label{fig:h2o} \end{figure}

For a non-linear three-atomic molecule such as water, many more symmetry-unique orientations 
of the external field exist than for linear molecules. Therefore, the present study has been 
restricted to four representative orientations: parallel (A) and antiparallel (B) to the molecular 
axis, perpendicular to the molecular axis in the molecular plane (C), and perpendicular to 
the molecular plane (D). Figure \ref{fig:h2o} shows Stark shifts, ionization rates, and second 
moments as a function of field strength for these four orientations. All quantities were calculated 
with HF and CCSD and the complex energy (that is, $\Delta E$ and $\Gamma$) was additionally 
evaluated at the CCSD(T) level of theory. 

The upper left panel of Figure \ref{fig:h2o} illustrates that electron correlation makes little 
impact on the Stark shift except for case B where the external field operates against the 
intrinsic dipole moment of the water molecule. Here, the effect of electron correlation is 
substantial in that CCSD(T) predicts zero Stark shift at around $F=0.13$ a.u., whereas that 
occurs at $F=0.16$ a.u. at the HF level. Likewise, a net dipole moment of zero (that is, a 
maximum in $\Delta E(F)$) is obtained at around $F=0.065$ a.u. with CCSD(T), but at around 
$F=0.078$ a.u. with HF. These discrepancies are somewhat unexpected given that CCSD(T) 
and HF agree within 7\% about the dipole moment of field-free H$_2$O. 
 
The lower panels of Figure \ref{fig:h2o} show that HF and CCSD(T) predict the same order 
of the orientations A--D regarding $\Gamma$: Ionization is at all considered field strengths 
easiest when the field is perpendicular to the molecular plane (case D) and hardest when the 
field is in plane perpendicular to the molecular axis (case C). The anisotropy parameter 
$\Gamma_D/\Gamma_C$ decreases with growing field strength from about 20 at $F=0.06$ 
to below 3 at $F=0.14$ a.u. These findings are in line with previous investigations using 
CAP-augmented TD-CIS.\cite{krause15b} Remarkably, $\Gamma$ is at all field strengths 
higher in case B where the field works against the intrinsic dipole moment than in case A 
where ionization occurs towards the electron-rich side of the molecule. This agrees with a 
previous investigation based on a single-electron Schr\"odinger equation.\cite{laso17} 

Even though electron correlation does not change the order of orientations A--D, it increases 
absolute values of $\Gamma$ considerably, in particular at low field strengths. At $F=0.06$ 
a.u., HF and CCSD ionization rates differ by a factor of 3--6 and the (T) correction additionally 
increases $\Gamma$ by up to 25\%. Those deviations shrink with rising field strength, but 
even at $F=0.14$ a.u., the difference between HF and CCSD is still about 30\% and that 
between CCSD and CCSD(T) about 10\%. 

From the upper right panel of Figure \ref{fig:h2o}, it is seen that in all four cases A--D, the 
component of the second moment in the direction of the field exhibits the characteristic peak 
discussed in Section \ref{sec:r2}, whereas the other components vary less with the field 
strength (see SI for details). Consistent with the highest ionization rates for orientation D, 
the peak is observed at the lowest field strength in this case. Eq. \eqref{eq:fabi} yields 
$F_\text{ABI}=0.057$ a.u. using the Stark-shifted lowest ionization potential (13.02 eV; 
12.67 eV at $F=0$) of H$_2$O, which is in very good agreement with the estimate based 
on the second moment. This suggests the transition from tunnel to barrier suppression 
ionization happens in analogy to a hypothetical atom with the same ionization potential 
for orientation D, while it is shifted to higher field strengths at the other orientations, 
most strongly in case C, which also features the lowest ionization rates.

\subsection{Methane}
\label{sec:ch4}
\begin{figure}[htbp]
\includegraphics[scale=0.895]{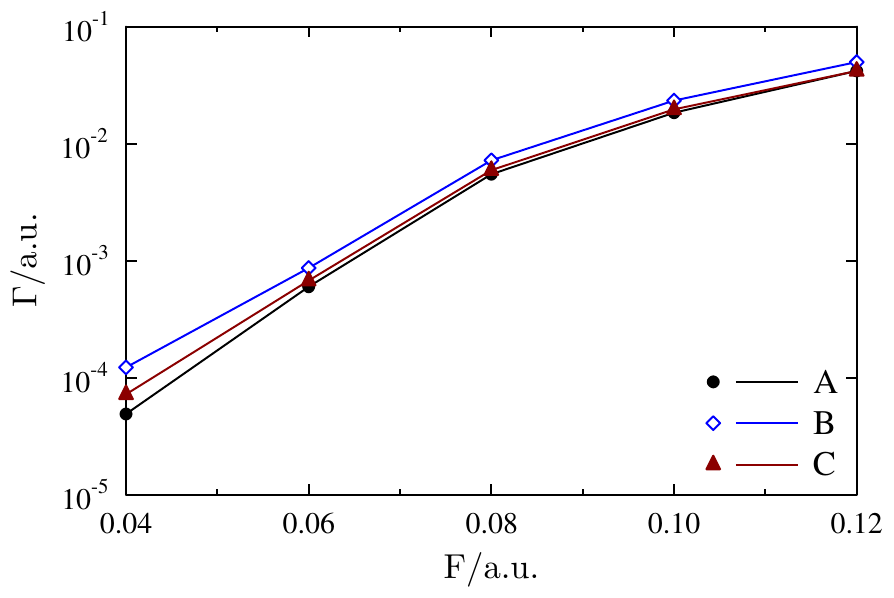}
\includegraphics[scale=0.895]{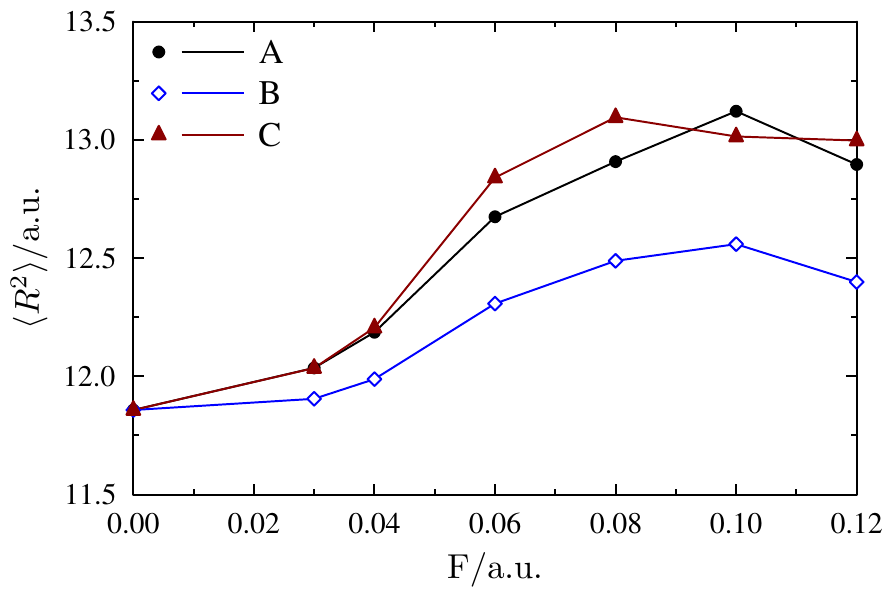} 
\caption{Strong-field ionization of CH$_4$ studied at the CCSD level of theory using a modified 
aug-cc-pVQZ basis set. Left panel: Ionization rates $\Gamma$ of CH$_4$ as a function of field 
strength. Right panel: Real part of the component of the second moment in the direction of the 
external field as a function of field strength. The field is oriented as follows: \\ 
A --- along a CH bond towards the hydrogen atom (= along the vector ($0 \; -\!\sqrt{2} \; -\!1$)), \\
B --- along a CH bond towards the carbon atom (= along the vector ($0 \;\; \sqrt{2} \;\; 1$)), \\
C --- bisecting the angle between two CH bonds (= along the $z$-axis)}
\label{fig:ch4} \end{figure}

Figure \ref{fig:ch4} shows CCSD ionization rates and second moments as a function of field strength 
for methane. Three orientations of the external field were studied: parallel (A) and antiparallel (B) 
to a CH bond and bisecting the angle between two CH bonds (C). As the left panel illustrates, the 
ionization rate depends only weakly on the orientation of the field consistent with the isotropic 
electron density distribution of methane. A slight preference exists for case B, where the electron 
leaves the molecule in the direction of a CH bond, but $\Gamma$ values corresponding to the three 
different orientations differ only by a factor of 1.2--1.5 in the range $F=0.06$--0.12 a.u (as opposed 
to up to 20 in the case of H$_2$O). HF underestimates $\Gamma$ by a factor of about 2 at F=0.06 
and by 10-20\% at $F=0.12$ a.u. (see SI for details), which is a relatively small deviation compared 
to H$_2$O and especially O$_2$. 

The right panel of Figure \ref{fig:ch4} demonstrates that the component of the second moment in the 
direction of the field shows the same characteristic behavior as for the other molecules discussed 
before. This suggests the onset of the barrier-suppression regime is at a field strength of around 
0.08 a.u. for orientation C and at around 0.10 a.u. for the other two orientations. For a hypothetical 
atom with the ionization potential of CH$_4$ (14.53 eV at $F=0.10$ a.u. and orientation C with 
EOM-IP-CCSD, 14.43 eV in the field-free case), that onset would be at 0.071 a.u. according to 
Eq. \eqref{eq:fabi}, that is, the tunnel ionization regime is only slightly extended compared to the 
atomic case.

\section{Conclusions and outlook}
\label{sec:conc}

This work has demonstrated that CCSD and CCSD(T) calculations in a basis set of atom-centered 
Gaussian functions with complex-scaled exponent deliver an accurate description of molecular 
strong-field ionization. Stark shifts and static-field ionization rates are computed as complex 
eigenvalues of the time-independent many-body Schr\"odinger equation without invoking any 
further approximation besides using a finite one-electron basis set and truncating the CC 
expansion. Key advantages of the proposed method are that it can be applied to polyatomic 
species with arbitrary molecular structure (subject to the usual constraints about the 
validity and applicability of CC methods \cite{ccbook}) and that the computation of molecular 
properties is straightforward. 

Results for the two-electron systems He and H$_2$ are in excellent agreement with reference 
values as long as the ionization rate is not smaller than ca. $10^{-5}$ a.u. Selected applications 
to many-electron systems illustrate the huge impact of electron correlation on the ionization rate, 
especially in the tunnel ionization regime. For example, HF and CCSD ionization rates of O$_2$ 
differ by a factor of up to 20 at low field strengths and the methods also disagree about the 
orientation of the external field at which the ionization rate is highest. For the polar molecules CO 
and H$_2$O, electron correlation also makes a significant impact on the Stark shift. 

A distinct maximum in the component of the second moment of the electronic charge distribution 
in the direction of the external field is observed for all considered molecules (H$_2$, O$_2$, 
H$_2$O, CH$_4$) at a certain field strength. This peak can be associated with the transition 
from tunnel to barrier suppression ionization by analogy with atomic Stark resonances. The 
position of the peak can vary substantially with the orientation; in particular, it is shifted 
to higher field strengths at orientations where ionization is suppressed. 

While the results reported here are very encouraging, it is also clear that further work along several 
lines is needed: First, because of the extreme basis-set requirements, the CC treatment presented 
in this work is suitable only for small molecules. This makes the implementation of more cost-effective 
methods desirable in order to study strong-field ionization of larger molecules. Second, the computation 
of partial widths of individual channels needs to be enabled for a more detailed characterization of the 
ionization process. Third, an extension of the current method to Stark resonances in time-dependent 
fields appears worthwhile to pursue. 

\section{Supplementary Material}

See supplementary material for details about the basis sets and molecular structures, numerical 
results corresponding to Figures \ref{fig:h2}--\ref{fig:ch4} and additional results.

\section{Acknowledgments}
This work has been supported by the Emmy Noether program of the Deutsche 
Forschungsgemeinschaft (grant JA 2794/1-1).

\clearpage

\section{Tables}

\begin{table}[htbp]
\caption{Estimates of the critical field strength at which the transition from tunnel to barrier 
suppression ionization occurs in the H$_2$ molecule.} \vspace{0.2cm}
\begin{tabular}{cccc} \hline
orientation & R(HH)/\AA & $I_p$/eV$^a$ & $F_\text{ABI}$/a.u.$^b$ \\ \hline
perpendicular & 0.74 & 17.00 & 0.098 \\
perpendicular & 1.40 & 13.73 & 0.064 \\
parallel & 0.74 & 17.14 & 0.095 \\
parallel & 1.40 & 13.70 & 0.058 \\
\hline
\end{tabular} \label{tab:h2fabi}
\footnotetext{Computed with EOMIP-CCSD at F=0.10 a.u. (R(HH)=0.74 \AA) and F=0.06 a.u.
(R(HH)=1.40 \AA).}
\footnotetext{Calculated according to Eq. \eqref{eq:fabi} for perpendicular orientation 
and using a double-well model for parallel orientation.}
\end{table}

\section{Figures}

\end{document}